\documentclass[final,3p,times]{elsarticle}

\usepackage{amssymb}
\usepackage{xcolor}
\usepackage{siunitx}
\usepackage{textgreek}

\journal{Elsevier}

\usepackage{subcaption}
\usepackage{graphicx}
\usepackage{float}

\begin{document}

\begin{frontmatter}



\title{Gravity and Composition Modulated Solidification and Mechanical Properties of $Al$-$Cu$ Nanostructures}


\author[label1]{Apurba Sarker} 

\author[label1]{Sourav Saha\corref{cor1}}
\ead{souravsaha@vt.edu}
\cortext[cor1]{Corresponding author}
\affiliation[label1]{organization={Kevin T. Crofton Department of Aerospace and Ocean Engineering, Virginia Polytechnic and State University},
            addressline={1600 Innovation Drive}, 
            city={Blacksburg},
            postcode={24060}, 
            state={VA},
            country={United States}}

\begin{abstract}
The future of space exploration and human settlement beyond Earth hinges on a deeper understanding of in-space manufacturing processes. The unique physical conditions and scarcity of experimental data demand robust computational models to investigate the atomic-scale physics of solidification. This work presents a molecular dynamics (MD) model to examine the solidification behavior of the $Al$–$Cu$ binary alloy, focusing on the influence of varying compositions and gravity levels (Earth, Lunar, Martian, and microgravity) on atomistic solidification mechanisms and the resulting mechanical properties—specifically, hardness—of \textit{as-solidified} nanostructures. Hardness is evaluated via nanoindentation simulations. The study confirms that gravitational forces significantly affect the solidification pathways of $Al$–$Cu$ alloys. Notably, by tuning alloy composition, the influence of gravity can be modulated—and in some cases, even reversed. Moreover, hardness exhibits a coupled dependence on both composition and gravity, offering a promising avenue for \textit{bottom-up} design of components tailored for extraterrestrial environments. The article delves into the nanoscale physical mechanisms underlying these phenomena and outlines future directions for extending this modeling framework to broader applications.     
\end{abstract}



\begin{keyword}
In-space Manufacturing \sep Solidification \sep Gravity \sep Molecular Dynamics \sep Nanoindentation 
\PACS 0000 \sep 1111
\MSC 0000 \sep 1111
\end{keyword}

\end{frontmatter}


\section{Introduction}\label{sec:Introduction}

In-space welding and manufacturing trace their origins to pioneering experiments in the 1960s \cite{sowards2023welding}. Today, the field is experiencing a resurgence, driven by the bold ambition of enabling sustained human exploration and eventual settlement on Mars \cite{mccoy2024beyond,subin2025space}. Electron-beam welding (EBW) was historically the first fusion-based technique tested in orbit, notably during Soyuz missions in the 1960s and 1970s \cite{sowards2023welding, bourgeois1973convection}. However, EBW has significant drawbacks in space environments: it requires vacuum chambers, is highly power-intensive, and produces narrow penetration profiles that limit scalability \cite{peebles2019review}. In contrast, laser-based welding and additive manufacturing methods have gained favor due to their higher efficiency, greater process flexibility, and ability to operate without bulky vacuum enclosures \cite{reitz_additive_2021, zocca_challenges_2022}. Manufacturing in space introduces a fundamentally different solidification environment due to the absence of gravity-driven phenomena such as buoyancy and convection \citep{o2025establishing}. While microgravity may offer some benefits such as reduced oxidation, lower heat loss, and deeper laser penetration during welding \citep{reitz_additive_2021} it also disrupts solidification pathways in ways that directly affect microstructure formation. The suppression of buoyancy in microgravity conditions leads to non-uniform melt pool dynamics, variable layer thickness, poor adhesion, and inconsistent feeding \citep{snyder_effects_2013}. In this regime, surface tension dominates, often causing molten materials to bead into spheres rather than form continuous tracks \citep{reitz_additive_2021}. For powder-based processes such as Laser Beam Melting, which rely on gravity for powder spreading and deposition, this absence of gravitational stabilization becomes especially problematic \citep{zocca_challenges_2022}. The resulting parts tend to display higher brittleness and reduced reliability compared to Earth-based counterparts \citep{ishfaq_opportunities_2022}. Furthermore, levitation of both liquids and powders in microgravity can contaminate the build and degrade layer quality \citep{ishfaq_opportunities_2022}. These interesting phenomena make investigating the solidification of alloys under reduced gravity an attractive proposition.

To date, additive manufacturing (AM) in space has been limited primarily to polymers (e.g., ABS) using simple extrusion-based systems such as 3DPrint and AMF aboard the International Space Station (ISS) \citep{sacco_additive_2019}. No significant studies have systematically investigated metallic solidification or composite processing in orbit, apart from early Skylab experiments \cite{bourgeois1973convection}. Because ISS testing is slow, costly, and constrained, parabolic flights, drop towers, and sounding rockets have been used to approximate the influence of microgravity from ground-based experiments. However, these methods only provide seconds to minutes of reduced gravity \cite{zocca_challenges_2022}, insufficient to capture the full evolution of solidification and microstructure.

As a result, simulation and mathematical modeling remain the most promising avenues for advancing space manufacturing, particularly at the length- and time-scale of solidification and microstructure formation. Contemporary modeling efforts on microstructure development at that length- and time-scales have spanned particle-level Discrete Element Method (DEM) \cite{steuben_discrete_2016}, mesoscale melt pool simulations \cite{korner_fundamental_2013}, macroscale thermal-mechanical models \cite{ninpetch_thermal_2019}, and atomistic or coarse-grained molecular dynamics for microstructure development \cite{kumar_micro-and_2018, tong_coarse-grained_2019, baghel_numerical_2022, guo_unraveling_2022}. Yet, the physics of solidification in microgravity, particularly the pathways that determine microstructural morphology, grain orientation, and defect formation, remains largely unexplored. Current understanding relies heavily on sparse experimental insights \cite{tong_molecular_2021}, underscoring the urgent need for rigorous solidification modeling frameworks that can predict microstructure evolution in extraterrestrial manufacturing environments. Prior molecular dynamics investigations have illustrated how nanoscale processing conditions influence microstructure and mechanical outcomes—for instance, examining the effects of cooling rate, dislocation dynamics, and voids on the deformation behavior of Inconel-718 nanostructures \cite{saha_inconel718_md_2021}, and exploring the role of filler geometry, volume fraction, and temperature in CNT–POM composites \cite{saha_w_nanocomposite_md_2020}.

There is strong experimental evidenue suggesting that the presence or absence of gravity fundamentally alters solidification dynamics. In microgravity, buoyancy and convection are suppressed, reducing solute transport and promoting irregular growth morphologies \citep{snyder_effects_2013, reitz_additive_2021}. As surface tension becomes dominant, molten droplets prefer spherical shapes, modifying the interfacial kinetics of advancing solidification fronts \citep{reitz_additive_2021}. Complementing these observations, molelar dynamics (MD) studies that emulate microgravity—e.g., by randomizing atom velocities or removing gravitational body-force terms in coupled fluid descriptions—reveal significant differences in atomic motion and microstructural evolution relative to terrestrial conditions \citep{tong_coarse-grained_2019, guo_unraveling_2022}. Consistent with these simulations, experiments tracking metallic solidification in reduced gravity report distinct microstructures compared to Earth-based cases \citep{tong_molecular_2021}. Together, these results indicate that gravity decisively reshapes solidification pathways, microstructure selection, and phase evolution.

If microstructure is altered by gravity, then mechanical behavior must also change. ISS-based AM studies report that parts printed under microgravity tend to be more brittle than their Earth-fabricated counterparts, consistent with modified solidification routes \citep{ishfaq_opportunities_2022}. Process signatures linked to reduced gravity—poor adhesion, unstable layering, and variable thickness—degrade structural continuity and amplify anisotropy \citep{snyder_effects_2013}. MD simulations of $Al$ solidification further support this structure–property connection: distinct nucleation modes (isothermal vs.\ quench) produce different grain morphologies that govern proxies for hardness and toughness \citep{mahata_understanding_2018}. Athermal heterogeneous nucleation of $Al$ on $Ti$ particles shows that subtle shifts in nucleation pathways (adsorption-driven vs.\ uniform growth) can markedly change cap morphology and bonding strength \citep{fujinaga_molecular_2019}. Thus, gravity-driven changes in solidification directly map onto mechanical property differences through their microstructural imprints.

$Al–Cu$ alloys have long been established as one of the most studied binary metallic systems in solidification science, serving as a canonical testbed for investigations of microstructure evolution \cite{haapalehto_atomistic_2022, mahata_understanding_2018}. Their binary phase diagram is relatively simple and well-characterized, enabling controlled studies of solute partitioning, dendritic growth, and eutectic transformation pathways \cite{li_performance_2015, baghel_numerical_2022}. Although commercial aerospace alloys incorporate multiple alloying additions, the binary $Al–Cu$ system isolates the primary solute interaction, making it especially tractable for molecular dynamics analysis and facilitating clearer connections between mechanistic insights and microstructural selection. Moreover, copper is a critical strengthener in aerospace-grade aluminum alloys, and variations in $Cu$ concentration directly influence hardness, dislocation density, and phase selection \cite{tong_molecular_2021, guo_unraveling_2022}. Consequently, $Al–Cu$ alloys offer both fundamental tractability and direct aerospace relevance, justifying their selection for systematic studies of solidification under reduced-gravity conditions.

For $Al–Cu$ alloys, solute concentration is a knob that tunes solidification and the resulting microstructure. Bond-order-potential MD simulations show that solute partitioning governs solid–liquid coexistence and interfacial kinetics \cite{haapalehto_atomistic_2022}, while parallelized MD with MEAM potentials indicates that composition modulates nucleation rates and selects distinct microstructural regimes that, in turn, control defect density and strength \cite{mahata_understanding_2018, li_performance_2015}. Across experiments and computations, $Cu$ consistently emerges as a microstructural control element: adjusting Cu content can steer phase selection, defect formation, and dislocation density—the root causes of hardness and brittleness shifts under both Earth gravity and microgravity. Therefore, targeted Cu tuning provides a practical pathway to counteract adverse mechanical outcomes induced by microgravity-altered solidification.

The evidence above motivates a solidification-structure-property nexus simulation framework that isolates gravity’s role in structural phase formation during solidification, explicitly links those mechanisms to mechanical properties, while using the composition as a controllable design variable. This article provides the first systematic all-atom molecular dynamics investigation, to the best knowledge of the authors, that analyzes the effect of reduced gravity levels that simulate specific environments (Mars (0.36g), Moon (0.16g), and micro ($10^{-6}$g)) on solidification behavior of $Al-Cu$ binary alloys. In addition, this work performs carefully designed numerical nano-indentation experiments on \textit{as-solidified} structures to determine the hardness as a representative mechanical property. The article is organized as follows: Section \ref{sec:Method} will describe the computational methodology and validation of the said methodology, Section \ref{sec:Analysis_and_discussion} provides evidence and detailed discussion on the effect of gravity and composition of $Al-Cu$ on the solidification behavior and mechanical properties. Section \ref{sec:Limitation} mentions the limitations and future extensions of this work, and Section \ref{sec:Conclusion} wraps up the article, outlining the key findings.

\section{Methodology} \label{sec:Method}
\subsection{Molecular Dynamics Simulation of Solidification}
\label{sec:Solidification}

This study employs classical MD simulations to investigate the directional solidification behavior of $Al–Cu$ binary alloys under varying gravitational conditions and alloy compositions. All simulations are conducted using the Large-scale Atomic/Molecular Massively Parallel Simulator (LAMMPS) framework \cite{plimpton_lammps_1995}. Interatomic interactions are described using the Embedded Atom Method (EAM), which captures metallic bonding through a combination of many-body embedding terms and pairwise potentials that account for electrostatic forces and short-range repulsion. For $Al–Cu$ systems, the potential developed by Cai and Ye \cite{cai_ye_eam_alcu_1996} is utilized, providing reliable thermodynamic fidelity and phase stability across the compositional range.

\begin{figure}
  \centering
  \includegraphics[width=0.85\textwidth]{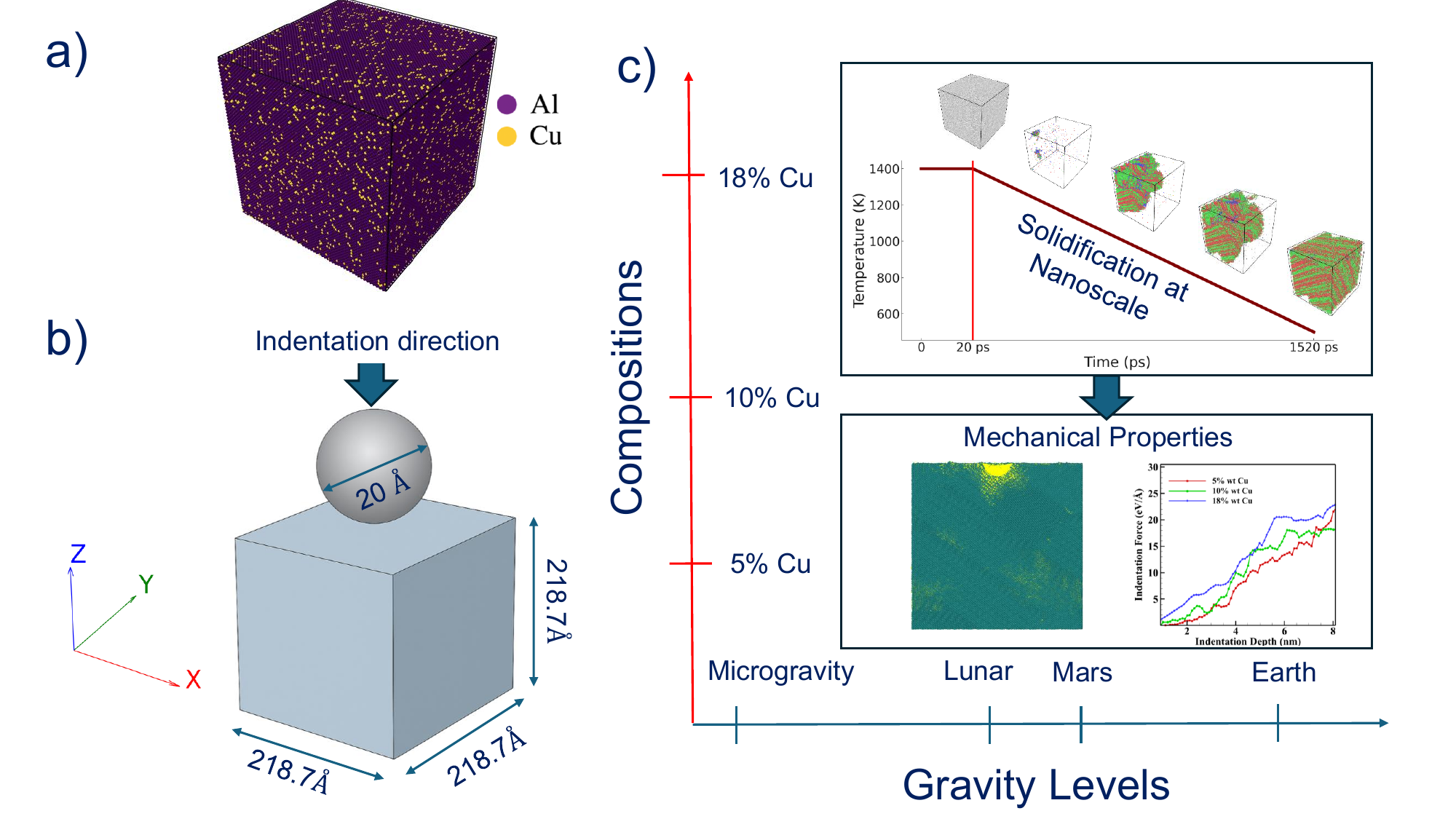}
  \caption{Methodology of solidification analysis and nanoindentation study under varying gravity levels and copper compositions. a) Solidification blocks of $Al–Cu$ bianry alloys with 5, 10, and 18 wt.\% $Cu$ are modeled to investigate the influence of copper concentration; b) nano-indentation simulations are performed using a spherical indenter of 20 \si{\angstrom} diameter on the solidified cubes; and c) the upper panel illustrates the temporal evolution of temperature and microstructure during solidification, while the lower panel presents the mechanical properties extracted from the nano-indentation simulations.}
  \label{fig:Fig1}
\end{figure}

The computational cell is a cubic domain with edge length 21.87 nm, representing a nanoscale alloy volume. Three alloy compositions are considered: 5 wt.\%, 10 wt.\%, and 18 wt.\% $Cu$ which are chosen because experimental microgravity solidification validation is available in literature for these alloys \cite{williams_beckermann_benchmark_2023}, providing a benchmark for comparison. Gravitational effects are introduced by applying a uniform body force along the Z axis downwards (0, 0, -1), with magnitudes corresponding to Earth gravity (1 g), Martian gravity (0.36 g), Lunar gravity (0.16 g), and an effective microgravity case ($10^{-6}$ g). These values are converted into the metal unit system of LAMMPS, then scaled appropriately to represent each gravity condition.

The simulation framework unfolds in two distinct stages, as illustrated in Figure \ref{fig:Fig1}a) and c). During the initial equilibration phase, the alloy system is uniformly heated to 1400 K—well above its melting point—to ensure complete liquefaction. The system is then evolved under isothermal–isobaric (NPT) conditions using a 2 fs timestep over a 20 ps duration, facilitating density relaxation and realistic atomic packing within the liquid phase. Gravitational acceleration is applied throughout this stage to capture its influence on thermal expansion and atomic distribution.

In the subsequent directional solidification stage, a controlled temperature gradient is imposed along the $X$ direction of the domain to initiate crystal growth. The box is divided into five zones: a cold end at 500 K, a hot end at 700 K, and three intermediate slabs that evolve freely, forming a continuous thermal gradient. Langevin thermostats are applied only to the hot and cold boundaries, while the interior evolves under microcanonical (NVE) dynamics, enabling a self-consistent propagation of the solid–liquid front. This stage runs for 1.5 ns with a 1 fs timestep. To reflect realistic conditions, lateral directions employ periodic boundary conditions, whereas the solidification axis is treated with free surfaces.

Each alloy composition is simulated across all four gravity levels, producing twelve distinct cases. During all simulations, atomic trajectories, thermodynamic properties, and spatial temperature distributions are monitored. Post-processing involves phase identification, dislocation density calculations, and interface morphology analysis. Finally, the fully solidified structures are subjected to nanoindentation simulations to quantify mechanical response (hardness) under different combinations of composition and gravity.

\subsection{Molecular Dynamics Simulation of Nanoindentation}
\label{sec:Nanoindentation}
The starting configurations for this stage are the solidified structures obtained after 1520 ps of directional solidification, previously simulated for three alloy compositions (5 wt.\%, 10 wt.\%, and 18 wt.\% $Cu$) under four different gravitational conditions (Earth, Martian, Lunar, and microgravity). These atomic configurations are imported directly into the nanoindentation simulation cell, where periodic boundaries are maintained in the lateral ($X$ and $Y$) directions and a free surface is kept along the $Z$ to allow realistic surface deformation. Atomic interactions are described by the same Embedded Atom Method (EAM) potential. To represent a rigid base, the atoms in the bottom 5~\AA\ of the sample are immobilized, while the remainder of the system is equilibrated at 300 K to remove any residual stresses from solidification. A rigid spherical indenter of 20~\AA\ diameter is then positioned just above the free surface and driven vertically into the material at a constant rate of $0.1~\text{\AA}/\text{ps}$ demonstrated in Figure \ref{fig:Fig1}b), applying a harmonic repulsive force to simulate contact mechanics without explicit modeling of the indenter’s atomic structure. The system evolves with a Langevin thermostat applied to dissipate heat generated during deformation. Throughout the simulation, the indentation depth, total potential energy, and vertical indentation force are recorded, and atomic configurations are saved periodically to track the development of plasticity and dislocation structures beneath the indenter. Each indentation proceeds to a maximum penetration depth of approximately 20 ~\AA\, enabling direct comparison of hardness and deformation mechanisms across alloy compositions and gravitational environments.

\subsection{Model Validation} \label{sec:Validation}
To ensure the correct and trustworthy model, the melting point of the alloy and lattice constant are determined and compared with literature. Determining the melting temperature for each $Al–Cu$ alloy composition provides a reference point for the simulation temperature scale. This calibration ensures that subsequent temperature settings can be interpreted consistently, allowing comparisons across different alloy compositions and with previous studies. This method further avoids artifacts associated with direct heating or cooling runs and yields a stable, reproducible estimate of the transition point.

\begin{figure}
  \centering
  \includegraphics[width=0.55\textwidth]{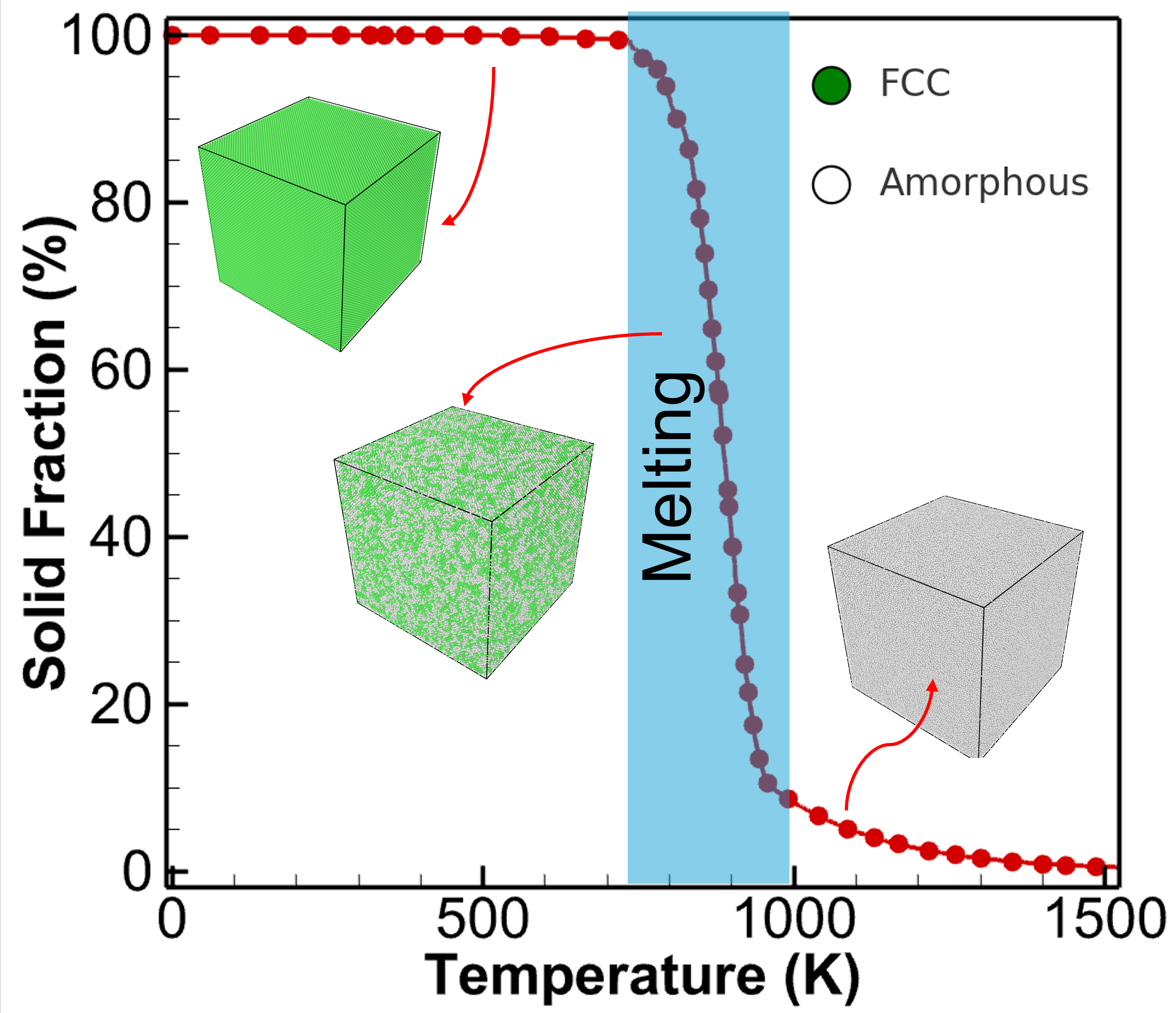}
  \caption{Evolution of solid fraction as a function of temperature during heating. It shows a sharp dip in solid fraction, indicating melting. Calculating threshold values of 90\% and 10\% of solid fraction, the melting range is calculated to be 816 K to 967 K.}
  \label{fig:melting-temperature}
\end{figure}

The melting phenomenon is obtained by monitoring the fraction of solid atoms within a central core region of the simulation cell, while excluding a 15 \si{\angstrom} boundary layer on all sides to minimize surface effects. The system is gradually heated under NPT conditions at $P=0$ bar, starting from 300 K and increasing to 1800 K in two stages: a rapid ramp to an intermediate temperature followed by a slower progression through the melting regime. Atomic order is identified using the centrosymmetry parameter \cite{PhysRevB.58.11085}, with atoms exceeding a threshold value (5.5) classified as liquid. The fraction of solid atoms in the core is then tracked as a function of temperature. The melting range is defined as $\Delta T_{90–10}$, corresponding to the temperatures where the solid fraction decreases from 90\% to 10\% \cite{Dai2024_MeltingPointMD} as shown in \ref{fig:melting-temperature}. For $Al$-5 wt.\% $Cu$, the solidus and liquidus limits occur near 816 K and 967 K, respectively. These results are consistent with earlier continuum and atomistic studies, which report melting intervals of 720–1120 K for 9 nm $Al–Cu$ nanoparticles \cite{Rahmani2018_CuAl_TiAl_MD, Puri2007_AlParticleSizeMelting, Huang2014_HollowCoreShellCollapse}, confirming that the present model reproduces realistic melting behavior. Sensitivity checks show that both $\Delta T_{10–90}$ and the transition temperatures are robust against moderate changes in the threshold criterion, with global solid fractions providing parallel confirmation of the results.

To establish that the simulated crystal is free of internal stress, the equilibrium lattice constant is determined at the target temperature and zero external pressure. Verifying this parameter ensures that the employed potential and alloy composition generate a physically realistic structure, free from artificial pressure, distorted packing, or biased elastic response. Such validation prevents errors from propagating into subsequent analyses of phase fractions, interface kinetics, and mechanical properties.

In practice, the equilibrium lattice constant is obtained by relaxing the simulation cell to zero hydrostatic pressure using a conjugate-gradient minimization scheme. Once minimized, the relaxed box dimension is divided by the number of FCC unit cells along each edge to compute the lattice parameter of the alloy:

\[ a_0 = 4.038\ \mathrm{\AA} \]

For the Al–5 wt.\% Cu supercell, this value aligns closely with the experimental X-ray diffraction measurement of 4.033 \si{\angstrom} reported by Zhang et al. \cite{Zhang2016_SciRep_31797}, differing by only 0.12\%. The agreement falls well within typical uncertainties of both empirical potentials and experimental measurements, confirming that the chosen interatomic potential provides a reliable structural baseline.

\section{Analysis and Discussion}\label{sec:Analysis_and_discussion}

\subsection{Effect of Reduced Gravity and $Cu$ Composition on Solid Fraction Evolution}

For every MD simulation snapshot, solid fraction is extracted using Common Neighbor Analysis (CNA) \cite{honeycutt1987cna}. CNA assigns every atom a local crystalline structure label (e.g., $\mathrm{fcc}$, $\mathrm{hcp}$, $\mathrm{bcc}$, $\mathrm{ico}$, or $\mathrm{other}$). The solid fraction is defined as the percentage of atoms identified as belonging to crystalline structures, relative to the total number of atoms in the system. For a frame at time $t$, let $N_{\alpha}(t)$ denote the number of atoms assigned label $\alpha$, and let $N(t)$ be the total atom count. The instantaneous solid fraction is then
\[
f_{s}(t)=\frac{\sum_{\alpha} N_{\alpha}(t)}{N(t)}.
\]

\begin{figure}[t]
  \centering  \includegraphics[width=0.75\textwidth]{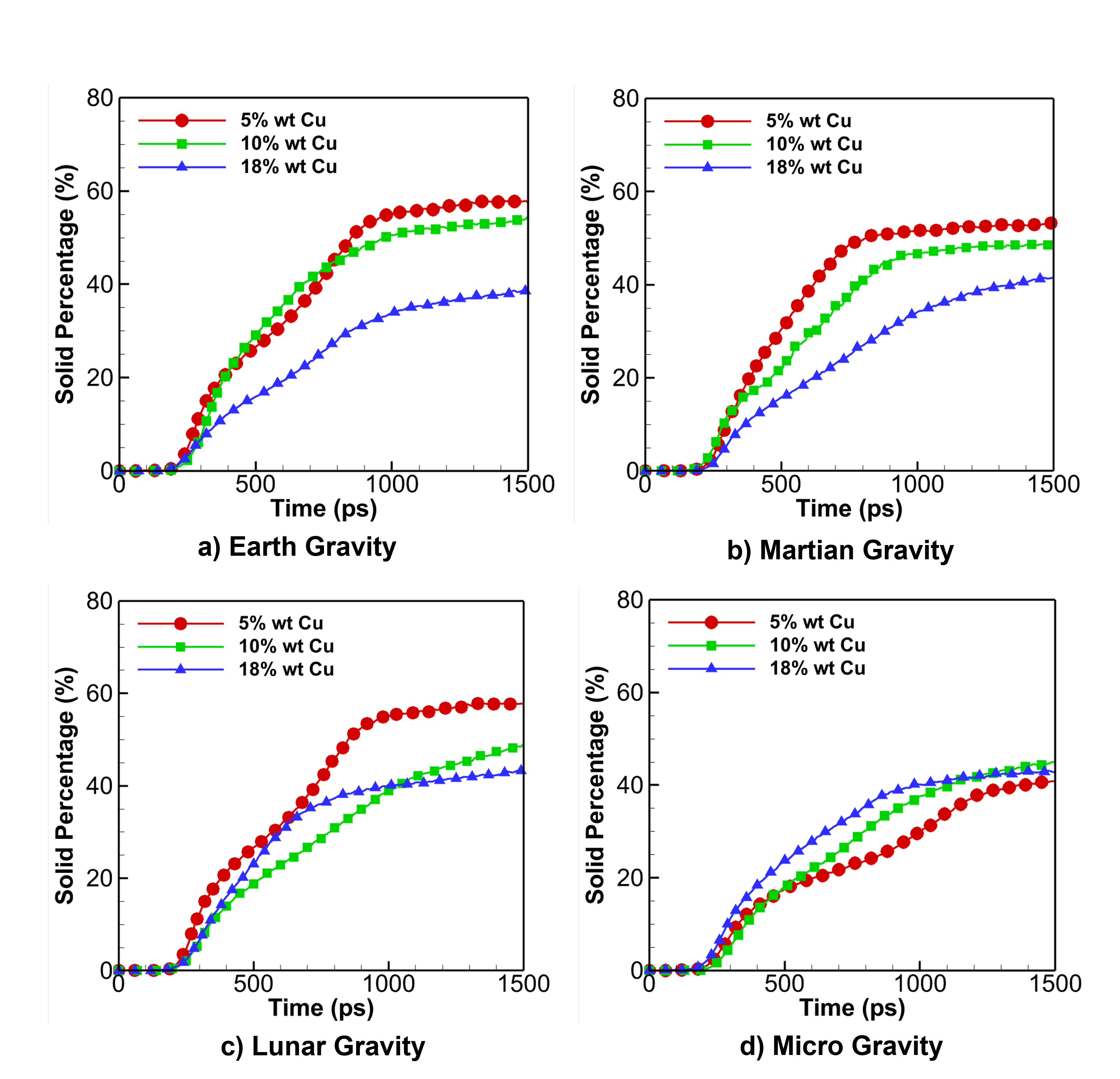}
  \caption{Solid fraction evolution for $Al–Cu$ alloys (5, 10, and 18 wt.\%) under different gravity conditions. Earth and Martian gravity accelerate solidification in dilute alloys, whereas microgravity reverses this trend, favoring FCC growth in $Cu$-rich systems.}
  \label{fig:solid-fraction}
\end{figure}

Firstly, for each gravity case, the effects of varying $Cu$ content are analysed. From Figure \ref{fig:solid-fraction}, it is observed that for both Earth gravity and Martian gravity, a lower $Cu$ content results in a higher rate of solidification. This is because the fraction of solid phase, that is predominantly FCC $Al$, decreases with increasing $Cu$ concentration in $Al–Cu$ alloys. This trend arises from the limited solubility of $Cu$ in the FCC $Al$ matrix and the thermodynamic preference for intermetallic formation at higher concentrations. Above the eutectic composition, the system favors nucleation of the tetragonal $\theta$-phase ($Al_2Cu$), which competes with FCC aluminum during solidification \cite{okamoto1990binary}. The presence of this second phase, combined with solute-induced undercooling, reduces the stability and growth of FCC domains.

Figure \ref{fig:solid-fraction} further shows the evolution of solid fraction for varying gravitational forces. Under Earth gravity, the 18~wt.\% $Cu$ alloy produces the lowest solid fraction, whereas the 5~wt.\% and 10~wt.\% systems solidify with progressively higher fractions. This ordering is consistent with prior atomistic simulations and experimental studies, which demonstrate that increasing $Cu$ concentration enhances solute disorder, frustrates FCC nucleation, and instead promotes the precipitation of the $\theta-Al_2Cu$ phase \cite{massalski_binary_1990, haapalehto_atomistic_2022}. A similar hierarchy persists under Martian gravity ($0.36g$), where the $Cu$-rich 18~wt.\% alloy continues to lag significantly behind the more dilute systems.

At Lunar gravity ($0.16g$), however, the picture begins to change. The gap in FCC fraction narrows, and the 18~wt.\% system starts to converge toward the lower-$Cu$ alloys. This marks the onset of a shift in the controlling physics: buoyancy-driven segregation of solute is weakened, so $Cu$ enrichment at the solid–liquid interface becomes less severe, and FCC fronts are less destabilized by local composition fluctuations.

In microgravity, the trend fully reverses. The 18~wt.\% alloy now produces the highest FCC fraction, surpassing both the 5~wt.\% and 10~wt.\% alloys. This reversal underscores the competing influences of solute segregation and interface stability. Under strong gravitational fields, $Cu$ atoms—being heavier—tend to settle toward the advancing interface. This segregation creates constitutional undercooling, destabilizes FCC growth, and promotes non-FCC morphologies such as $\theta$-$Al$$_2$$Cu$ networks \cite{mahata_understanding_2018, gu_computational_2019}. By contrast, in microgravity, the absence of buoyancy suppresses solute pile-up and promotes a more uniform chemical distribution. The resulting smoother and more stable solid–liquid interface enables efficient FCC ordering, even in alloys with high $Cu$ content. Thus, the relative ranking of compositions with respect to FCC fraction is inverted between terrestrial and microgravity conditions—a signature of the profound role that gravity plays in coupling solute transport to solidification dynamics.

Experimental studies support this observation as well. Solidification in microgravity produces planar or equiaxed fronts with lower dislocation densities and reduced solute trapping, conditions favorable to FCC growth \cite{zhang_comparative_2024, regel_improved_1999}. These results validate the simulated gravity-dependent reversal and demonstrate the coupled effect of alloy composition and gravitational field on nanoscale solidification pathways. However, this article investigates and confirms the underlying mechanisms at the nanoscale.

\begin{figure}
  \centering
  \includegraphics[width=1.0\textwidth]{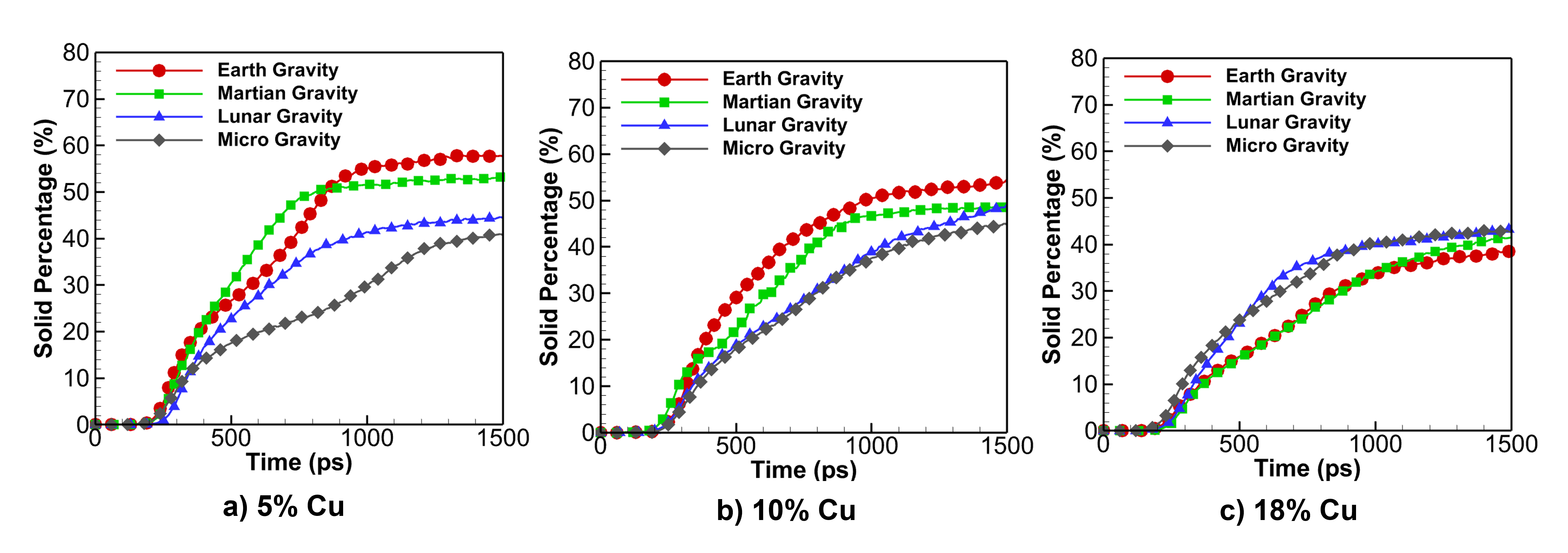}
  \caption{Solid fraction evolution at constant $Cu$ contents under different gravitational fields. Dilute alloys solidify more readily at high gravity, whereas $Cu$-rich alloys solidify more effectively under reduced gravity.}
  \label{fig:microstructure}
\end{figure}

Figure \ref{fig:microstructure} shows the effect of gravity level at constant copper content. For the 5~wt.\% alloy, Earth gravity produces the highest solid fraction, which steadily declines as gravity is reduced toward the microgravity regime. In contrast, the 18wt.\% alloy exhibits the opposite behavior: its FCC fraction is lowest on Earth but increases sharply under reduced gravity, reaching a maximum in microgravity. The 10~wt.\% case lies between these two extremes, showing only modest sensitivity to gravity level.

This crossover reflects the composition–gravity coupling outlined earlier. In dilute alloys, stronger gravity helps maintain interface stability, allowing FCC growth to dominate. In $Cu$-rich alloys, however, gravity intensifies solute pile-up at the front and destabilizes FCC order; only under weak or absent gravity does the interface recover enough stability for FCC solidification to proceed efficiently.

\subsection{Effect of Reduced Gravity and $Cu$ Composition on Microstructures}

\begin{figure}[h!]
  \centering
  \includegraphics[width=0.8\textwidth]{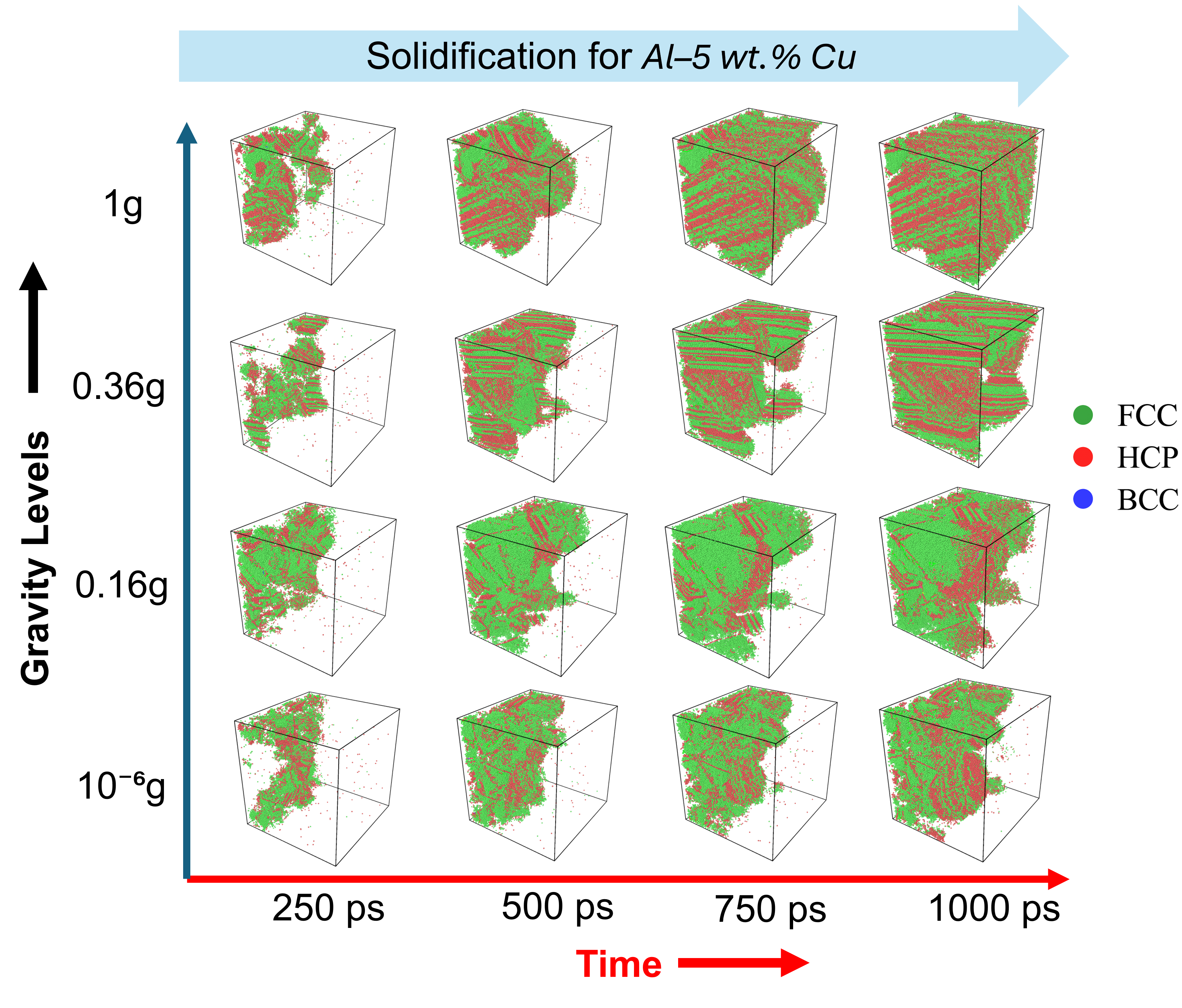}
  \caption{Microstructure evolution of $Al$–5 wt.\% $Cu$ alloy under Earth, Martian, Lunar, and microgravity at different times. Gravity accelerates nucleation and growth, while microgravity delays solidification and produces more uniform structures.}
  \label{fig:const_cu_solid_percentage}
\end{figure}

Figure \ref{fig:const_cu_solid_percentage} shows microstructure evolution during solidification of $Al$-5 wt\% $Cu$ at different gravity levels. It is observed that HCP phase also exists along with FCC even though pure aluminum is intrinsically FCC. The solidified $Al–Cu$ alloys consistently show a mixture of FCC and HCP atoms. This arises because the addition of $Cu$ alters the local atomic environment and promotes stacking irregularities during growth. FCC and HCP are both close-packed structures, differing only in their stacking sequence, so small perturbations in solute distribution or interface stability can drive the system to adopt HCP stacking locally. In concentrated or rapidly solidified alloys, these regions are stabilized as extended domains or defects, leaving behind HCP alongside the dominant FCC matrix.

The figure further reveals a general trend of earlier solidification onset with increasing gravitational force. In molecular dynamics simulations, gravity is applied as a body force, imparting a net downward pull that enhances atomic packing and stratification in the melt. This gravitational bias accelerates the migration of atoms toward energetically favorable lattice positions, thereby promoting earlier nucleation of the solid phase. Furthermore, gravity-driven settling facilitates the dissipation of latent heat released during solidification, effectively mimicking the role of buoyancy-driven convection observed in macroscopic systems. By contrast, in reduced or microgravity environments, the absence of such gravitationally induced rearrangements limits heat transport and reduces atomic mobility near the solid–liquid interface. This delays the development of sufficient under-cooling for nucleation, thereby postponing crystallization. Overall, gravity accelerates solidification kinetics by enhancing local atomic ordering and improving thermal energy dissipation. Hence, for Earth gravity, faster solidification is observed, which in turn decreases as the gravity is decreased. 

These observations align with prior experimental findings. Williams and Beckermann \cite{williams_beckermann_benchmark_2023} reported that $Al$–4 wt.\% $Cu$ alloys solidified under microgravity exhibited fully equiaxed grains, whereas terrestrial samples showed columnar-to-equiaxed transitions due to melt convection. Similarly, Zhang et al. \cite{zhang_comparative_2024} demonstrated that $Al$–10 wt.\% $Cu$ alloys form planar or equiaxed fronts in microgravity, while strong buoyancy-driven flow under 1 g conditions produces columnar and cellular morphologies. These results reinforce the critical role of gravity in controlling solidification pathways in $Al–Cu$ alloys.

\begin{figure}
  \centering
  \includegraphics[width=0.8\textwidth]{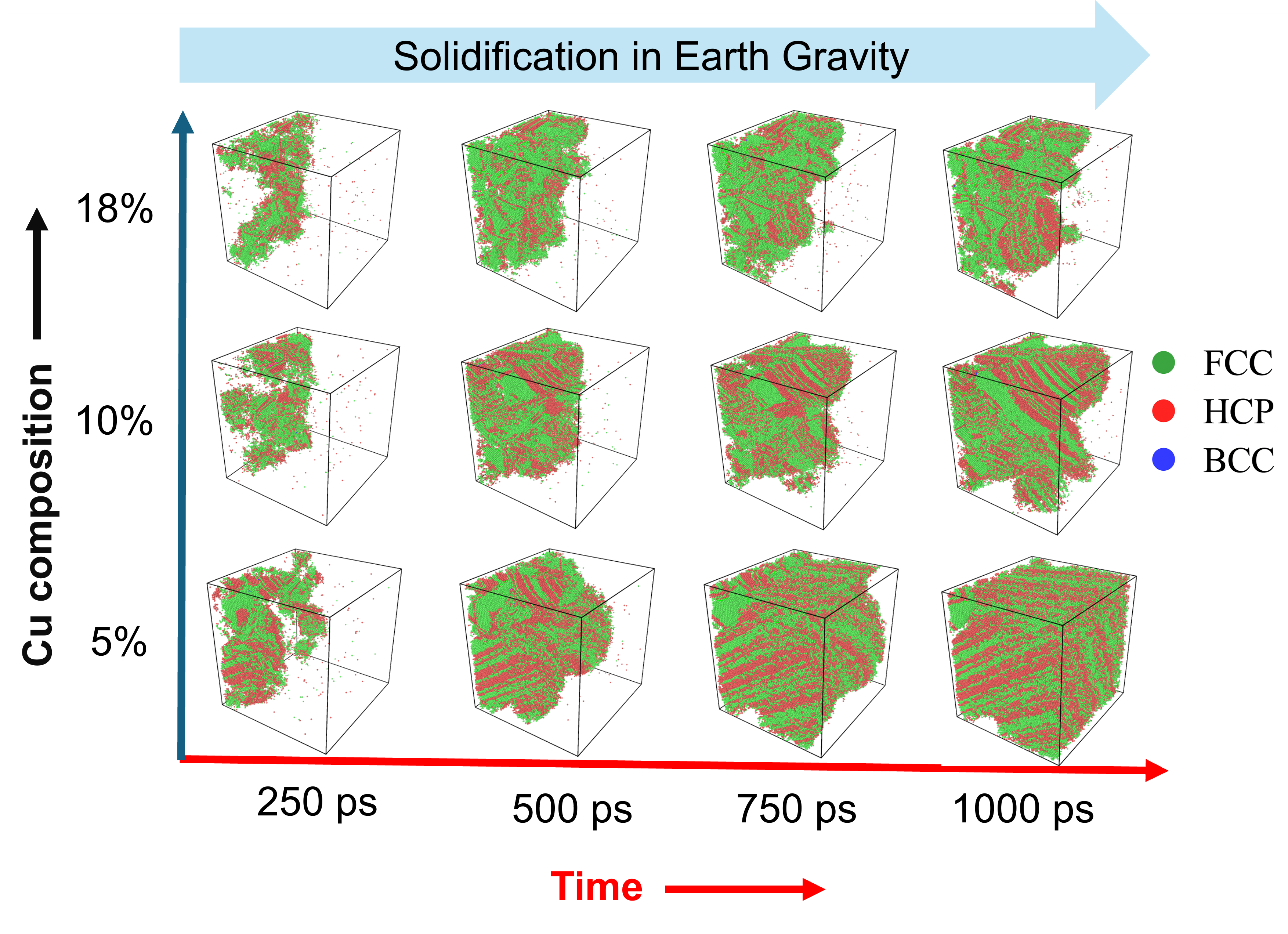}
  \caption{Microstructure evolution of $Al–Cu$ alloys under Earth gravity. Dilute alloys solidify faster and more directionally, while $Cu$-rich systems develop slower fronts and stacking irregularities.}
  \label{fig:microstructure_earth}
\end{figure}

\begin{figure}
  \centering
  \includegraphics[width=0.8\textwidth]{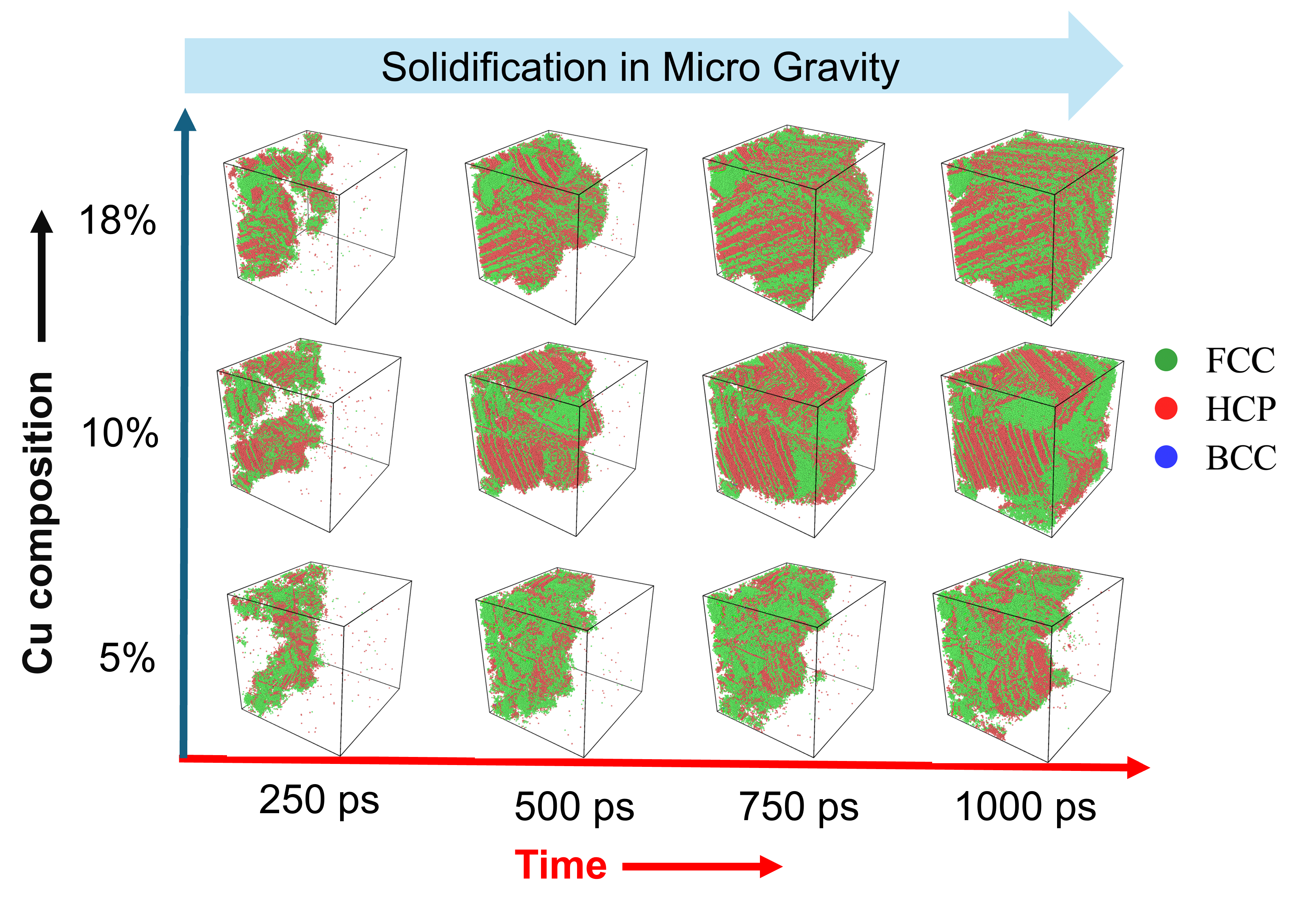}
  \caption{Microstructure evolution of $Al–Cu$ alloys under microgravity. High-$Cu$ alloys solidify more efficiently under reduced gravity due to uniform solute distribution and stabilized fronts.}
  \label{fig:microstructure_micro}
\end{figure}

Figures \ref{fig:microstructure_earth} and \ref{fig:microstructure_micro} illustrate the temporal evolution of microstructure for different $Cu$ compositions at Earth gravity and microgravity, respectively. Across both cases, the snapshots at 250, 500, 750, and 1000 ps clearly demonstrate the progression of directional solidification fronts, revealing how both gravity and solute concentration dictate the pace and morphology of growth.

Under Earth gravity (Figure \ref{fig:microstructure_earth}), the solidification front advances rapidly in the $5$ wt.\% $Cu$ alloy, producing extended FCC regions with relatively sharp directional order by 1000 ps. As $Cu$ concentration increases to 10 wt.\% and 18 wt.\%, the advance of the front slows down, and microstructures exhibit increased disorder, with more frequent HCP pockets arising from stacking irregularities. This delay results from gravity-driven $Cu$ segregation at the interface, which destabilizes FCC ordering and promotes local solute trapping. The trend is consistent with the solid-fraction results, but here the atomic-scale images emphasize the microstructural manifestation: less progress in directional growth at higher $Cu$ contents, accompanied by solute-induced irregularities.

In contrast, under microgravity (Figure \ref{fig:microstructure_micro}), the time-lapse reveals an inverted trend. The 18 wt.\% $Cu$ alloy solidifies more uniformly and develops extended FCC domains by 1000 ps, while the 5 wt.\% alloy shows a relatively slower progression of the front. Without buoyancy-driven segregation, $Cu$ atoms remain more evenly distributed, stabilizing the advancing interface and enabling efficient FCC ordering even at high concentrations. The microgravity case also displays fewer extended HCP regions, suggesting that the absence of gravity reduces solute pile-up and stacking faults.

Taken together, the two figures highlight that gravity not only changes the rate of solidification but also the pathway of microstructural evolution. Earth gravity accentuates solute segregation and stacking irregularities, limiting the advance of the growth front at higher $Cu$ contents. Microgravity, by contrast, suppresses segregation, leading to smoother fronts and faster directional progress at the same high-$Cu$ compositions. 

\subsection{Evolution of Dislocation Density during Solidification}\label{sec:Dislocation}

Dislocations are line defects within the crystal lattice that enable atomic planes to slip past one another under applied stress \cite{anderson2017theory}. The density of dislocations within a material strongly influences its mechanical response. A high dislocation density indicates a large number of such defects per unit volume. Under these conditions, dislocations interact, entangle, and obstruct each other’s motion, thereby increasing the resistance to plastic flow. This results in higher strength and hardness but simultaneously reduces ductility, making the material more prone to brittle failure \cite{gong2025multi}. Conversely, a low dislocation density corresponds to a relatively defect-free crystal lattice. With fewer barriers to dislocation motion, plastic deformation proceeds more readily, giving the material greater ductility and softness but lower yield strength. Thus, the interplay between strength and ductility is closely governed by dislocation density \cite{sadeghi2024dislocation}. Quantitatively, the dislocation density ($\rho$) is defined as:

\begin{center}
  \text{Dislocation density, $\displaystyle \rho = \frac{L_{\mathrm{total}}}{V}$}\\[4pt]
\end{center}

\(L_{\text{total}}\) is the total length of all dislocations, and $V$ is the volume of the material. Higher $\rho$ reflects greater defect concentration and enhanced hardness, while lower $\rho$ indicates fewer lattice imperfections and improved ductility.

\begin{figure}[H]
\centering
\includegraphics[width=0.85\textwidth]{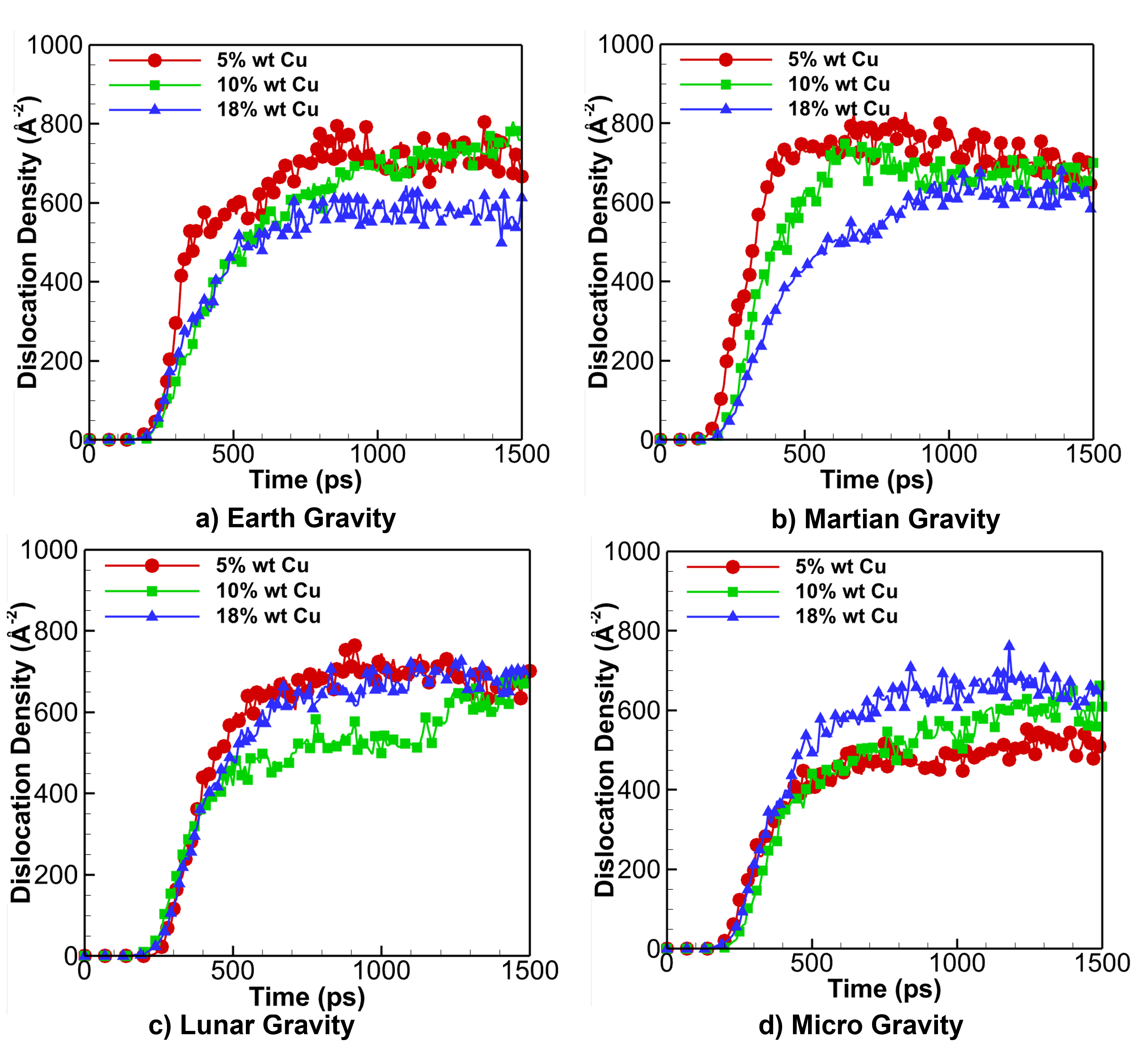}
\caption{Dislocation density evolution for $Al–Cu$ alloys at different gravity levels: a) Earth, b) Martian, c) Lunar, and d) micro gravity. Gravity enhances dislocations in dilute alloys, while microgravity promotes dislocations in $Cu$-rich alloys due to uniform solute distribution and lattice strain.}
\label{fig:Dislocation_at_const_gravity}
\end{figure}

In this study, dislocation density is found to vary strongly with both $Cu$ content and gravitational field. As demonstrated in Figure \ref{fig:Dislocation_at_const_gravity}, under Earth gravity, increasing the $Cu$ concentration from 5 wt.\% to 18 wt.\% leads to a marked decrease in dislocation density, from approximately 784~\AA$^{-2}$ to 630~\AA$^{-2}$ after 1500 ps. This reduction is attributed to solute-induced lattice strain and gravitationally driven $Cu$ segregation at the solid–liquid interface, which destabilizes FCC growth and limits the number of dislocations formed within the $Al$ matrix. These findings are consistent with prior molecular dynamics studies reporting that higher $Cu$ levels suppress defect-free FCC nucleation and promote structurally disordered morphologies \cite{haapalehto_atomistic_2022, mahata_understanding_2018}. In contrast, under microgravity, the trend reverses. The absence of buoyancy-driven convection and gravitational settling promotes more uniform solute distribution and smoother interface propagation, conditions that favor FCC nucleation even at high $Cu$ contents. As a result, dislocation density increases with $Cu$ concentration, with the 18 wt.\% alloy reaching 600 Å$^{-2}$ in microgravity, significantly higher than its Earth-gravity counterpart. These results align with experimental findings that microgravity solidification produces more equiaxed, defect-minimized microstructures with reduced segregation \cite{zhang_comparative_2024, regel_improved_1999}. Taken together, Figure \ref{fig:Dislocation_at_const_gravity} establishes that the role of $Cu$ is gravity-dependent: under strong gravity, it suppresses dislocation formation, while in microgravity it promotes it.

\begin{figure}[H]
\centering
\includegraphics[width=1.0\textwidth]{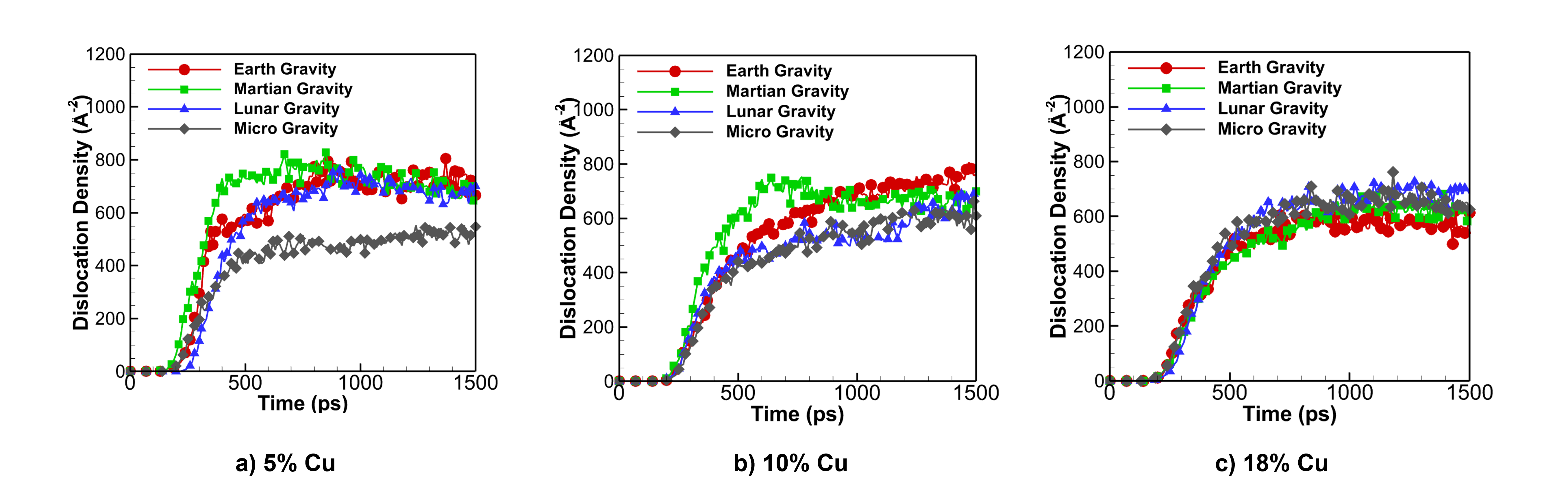}
\caption{Variation of dislocation density with gravity level for $Al–Cu$ alloys for a) 5\%, b) 10\%, and c) 18\% $Cu$ content. Dilute alloys generate more dislocations at high gravity, while $Cu$-rich alloys accumulate more defects under microgravity.}
\label{fig:Dislocation_at_const_Cu}
\end{figure}

The gravitational effect becomes even clearer when examined at constant composition, as shown in Figure \ref{fig:Dislocation_at_const_gravity}. For dilute alloys (5–10 wt.\% $Cu$), dislocation density increases with gravity because gravitational settling accelerates lattice ordering but also introduces mismatch stresses that seed additional defects. As $Cu$ content rises, however, gravity intensifies solute segregation and promotes the formation of $\theta$ ($Al_{2}Cu$) intermetallics. These ordered phases reduce the FCC volume fraction and the number of available slip systems, thereby lowering dislocation density in high-$Cu$ alloys. In microgravity, by contrast, solute atoms remain uniformly distributed, stabilizing FCC growth and preserving more slip systems. With a larger FCC volume available, more dislocations can nucleate and multiply, leading to higher dislocation densities in $Cu$-rich alloys than under Earth conditions. Thus, Figure \ref{fig:Dislocation_at_const_Cu} complements Figure \ref{fig:Dislocation_at_const_gravity} by showing that at low $Cu$ contents gravity enhances dislocations, whereas at high $Cu$ contents microgravity favors them. Together, these results highlight the coupled role of composition and gravity in setting the initial defect population inherited from solidification.

Figures \ref{fig:Dislocation_at_const_gravity} and Figure \ref{fig:Dislocation_at_const_Cu} show dislocation density is higher at higher gravity levels. In microgravity, the absence of significant gravitational pull leads to uniform atomic packing, minimal stress, and very few dislocations, resulting in low dislocation density. As gravity increases, atoms experience stronger downward forces. This creates stress gradients, promotes solute segregation, and disturbs solidification symmetry, leading to more dislocation nucleation. Thus, dislocation density increases with gravity. Previous experimental studies demonstrated that in detached (microgravity) directional solidification, ingots grow with little wall contact and exhibit dramatically reduced dislocation densities \cite{regel_improved_1999}.

\color{black}
\subsection{Nanoindentation under Different Gravity Levels}
\label{sec:nano_indent_gravity}

\subsubsection{Effects of Reduced Gravity and $Cu$ Composition on Hardness}
Nanoindentation is performed on the final structure received from the solidification simulations to study the mechanical property, hardness, as the gravitational force and composition change. From the results it is observed that under microgravity, the indentation behavior is more symmetric and relaxed, with lower dislocation activity and reduced hardness. As gravity increases, atoms beneath the surface experience a preloading effect that enhances resistance to deformation, resulting in higher indentation forces, increased defect generation, and greater hardness compared to microgravity. However, when $Cu$ content is varied under Earth gravity, both hardness and dislocation density decrease with rising concentration. At 5 wt.\% $Cu$, dislocation nucleation is most active, leading to higher hardness and stronger resistance to penetration. As the $Cu$ content increases to 10 and 18 wt.\%, solute segregation at the solid–liquid interface becomes more pronounced, stabilizing the interface and reducing defect generation. This segregation suppresses dislocation activity and lowers hardness, despite the intrinsic lattice strain introduced by additional solute atoms. Thus, the combined effect of gravity-driven segregation and solute stabilization causes both dislocation density and hardness to diminish with increasing $Cu$. Previous studies \cite{Bansal2011_AlCu_nanohardness} reported hardness of 0.63–0.75 GPa for $Al–4 wt.\% Cu$ nanocrystals, while a 2024 study on $Al–Cu$ MEMS films reported $Al$ film hardness of 0.4–0.8 GPa and $Al–0.5 wt.\% Cu$ hardness of 0.17–0.25 GPa \cite{Zhou2024_AlCu_MEMS}. The current study finds $Al–5 wt.\% Cu$ to have a hardness of 0.7316 GPa at plateau, consistent with these reported values and supporting the observed decreasing trend with higher $Cu$ contents. The elevated hardness value can be attributed to the size scale considered for simulation.

In microgravity, the trend of $Cu$ content versus dislocation density and hardness reverses compared to Earth, because the role of gravitational segregation and convection is suppressed. On Earth, higher $Cu$ promotes solute segregation at the solid–liquid interface, generating local stress fields and defect sources that multiply dislocations. In microgravity, however, buoyancy-driven convection is absent, so solute atoms distribute more uniformly in the melt. This reduces the extent of solute pile-up and interface instability that normally drives dislocation generation at higher $Cu$ contents. Instead, the dominant effect becomes solid-solution strengthening: at lower $Cu$ levels (e.g., 5 wt.\%), the solute is well dissolved and interacts strongly with dislocations, raising their density and hardness. As $Cu$ content increases to 10–18 wt.\% under microgravity, the lack of convection causes slower solute diffusion and the formation of more uniform but defect-suppressed microstructures, so dislocation density no longer rises and can even decrease. Thus, while Earth gravity amplifies solute segregation and defect nucleation with $Cu$, microgravity smooths solute distribution and stabilizes the interface, inverting the trend.

\begin{figure}[H]
  \centering
  \includegraphics[width=1\textwidth]{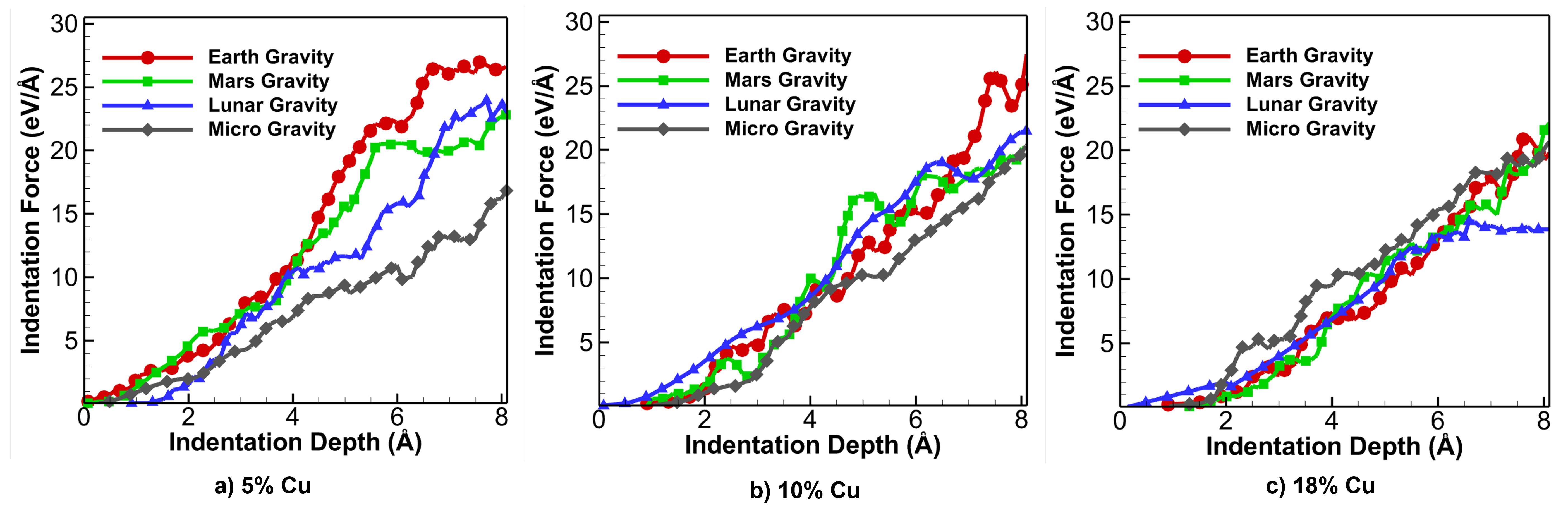}
  \caption{Nanoindentation force–depth curves for $Al–Cu$ alloys with different $Cu$ contents (a) 5\%, b) 10\%, and c) 18\%) under varying gravity levels. Gravity strongly enhances indentation resistance in dilute alloys, but its effect is reduced at higher $Cu$ contents.}
  \label{fig:11}
\end{figure}

Figure~\ref{fig:11} presents the indentation force--depth curves obtained from molecular dynamics simulations where a rigid spherical indenter of 20~\AA{} diameter is driven into the alloy surface at a rate of 0.1~\AA/ps. The bottom 5~\AA{} of the system is fixed to mimic a rigid substrate, and a Langevin thermostat dissipates local heating, ensuring a consistent comparison across cases. For the 5 wt.\% $Cu$ alloy, the indentation curves diverge strongly with gravity: microgravity produces a soft response, whereas Earth gravity exhibits a sharp rise in force due to enhanced dislocation nucleation and preloading. For 10 wt.\% $Cu$, the curves begin to converge, and at 18 wt.\% $Cu$, gravity’s influence is nearly eliminated as solute segregation dominates interface stability. This systematic weakening of gravity’s role with increasing $Cu$ is consistent with experimental nanoindentation studies, where dilute $Al–Cu$ alloys harden significantly while high-$Cu$ alloys show reduced sensitivity to external fields \cite{Bansal2011_AlCu_nanohardness,Zhou2024_AlCu_MEMS}.

\begin{figure}[H]
  \centering
  \includegraphics[width=0.7\textwidth]{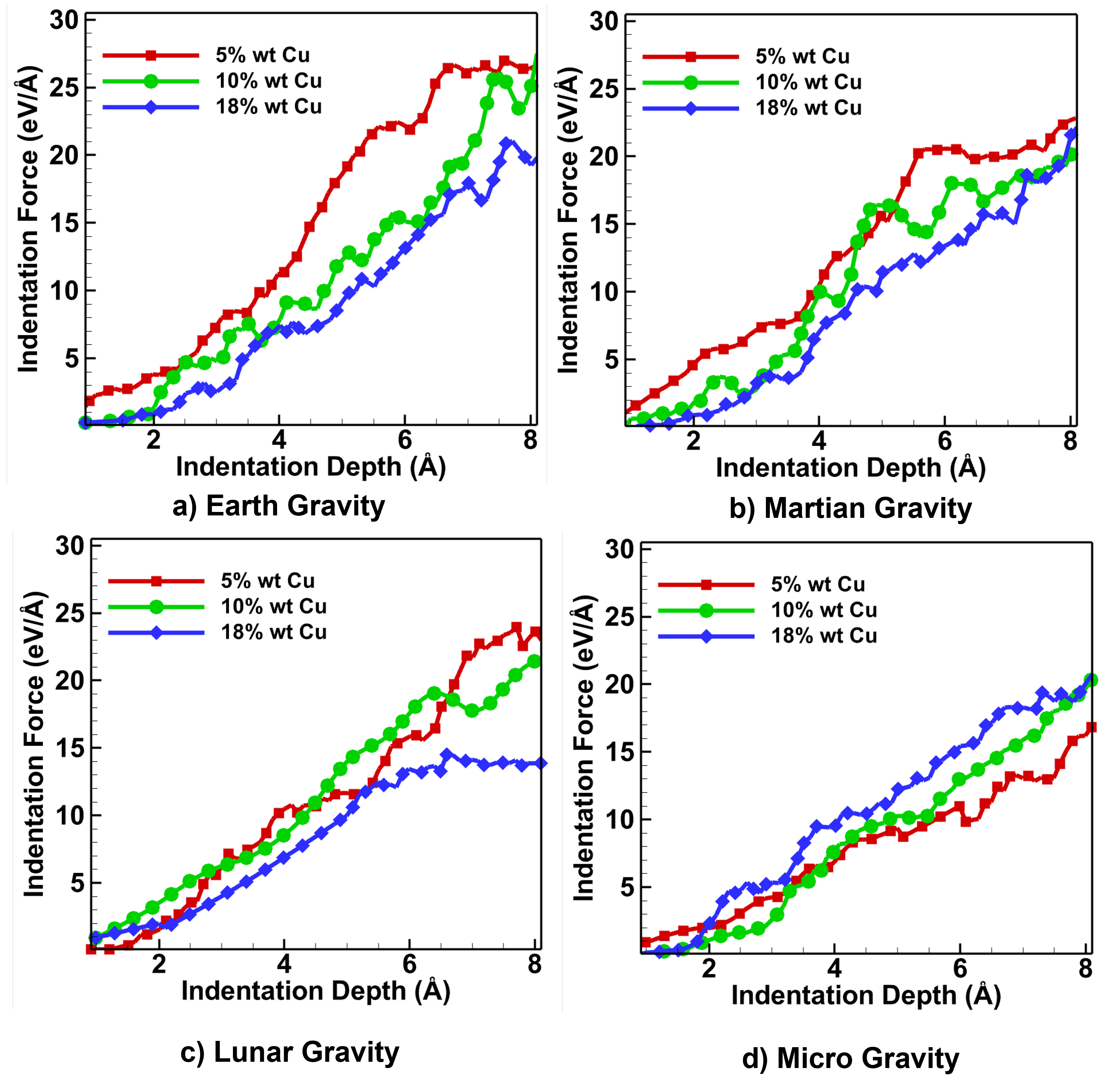}
  \caption{Nanoindentation response of $Al–Cu$ alloys under different gravity conditions with varying $Cu$ contents. a) Earth, b) Martian, c) Lunar, and d) Micro gravity. Earth gravity shows decreasing hardness with increasing $Cu$, while microgravity exhibits the opposite trend due to suppression of segregation and enhanced solid-solution strengthening.}
  \label{fig:12}
\end{figure}

Figure~\ref{fig:12} reorganizes the data by keeping gravity constant and comparing the role of $Cu$ concentration. Under Earth gravity, indentation resistance decreases with increasing $Cu$: the 5 wt.\% alloy shows the strongest force response, while the 18 wt.\% alloy is significantly softer. At Martian and Lunar gravities, the differences between compositions diminish, as segregation effects are weaker than at 1g. In sharp contrast, the microgravity case reveals an inversion of the trend—indentation force increases with $Cu$ content, with 18 wt.\% $Cu$ exhibiting the strongest resistance. The absence of buoyancy-driven segregation under reduced gravity ensures a more uniform solute distribution, allowing copper atoms to act as solid-solution strengtheners that resist slip and enhance lattice stability.  

The robustness of these observations is reinforced by methodological consistency across all cases: identical system sizes (21.87~nm cubic domain), indenter radius, maximum penetration depth ($\sim$20~\AA), and thermostatting protocols. Moreover, the simulated hardness values align with experimental ranges for nanocrystalline and thin-film $Al–Cu$ alloys \cite{Bansal2011_AlCu_nanohardness,Zhou2024_AlCu_MEMS}, supporting the physical fidelity of the force-depth curves. These findings demonstrate that the governing mechanism of hardness changes from gravity-driven segregation on Earth to solid-solution strengthening in microgravity, thereby capturing the fundamental inversion of trends across environments.

\begin{figure} [t]
  \centering
  \includegraphics[width=0.8\textwidth]{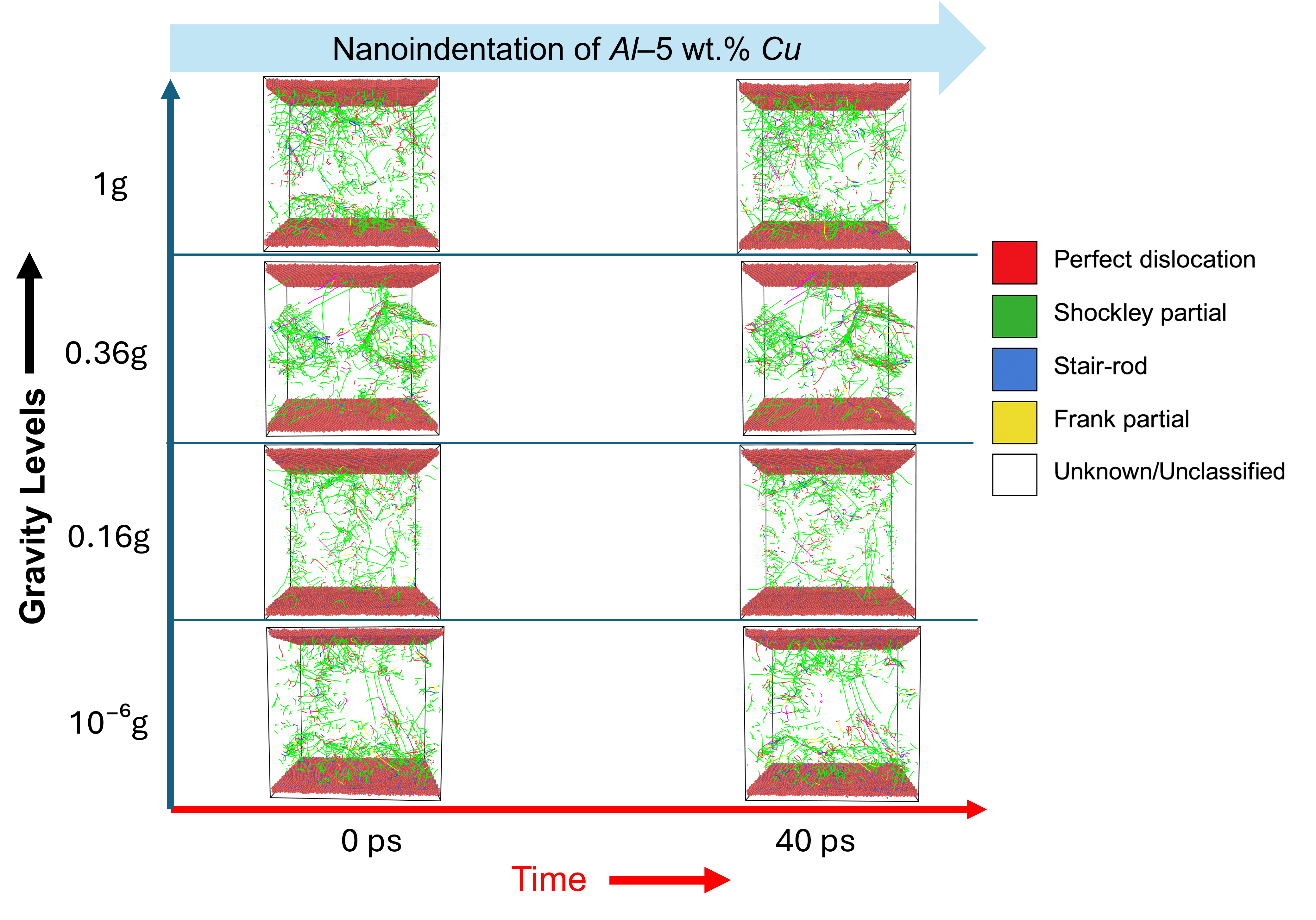}
  \caption{Dislocation structures' evolution during nanoindentation of Al–5 wt.\% $Cu$ alloy under varying gravity levels. Higher gravity enhances dislocation nucleation and multiplication, while microgravity results in fewer dislocations due to the defect-free solidification state.}
  \label{fig:13}
\end{figure}

\subsubsection{Effects of Reduced Gravity and $Cu$ Composition on Dislocation Density}
\begin{figure}
  \centering
  \includegraphics[width=0.8\textwidth]{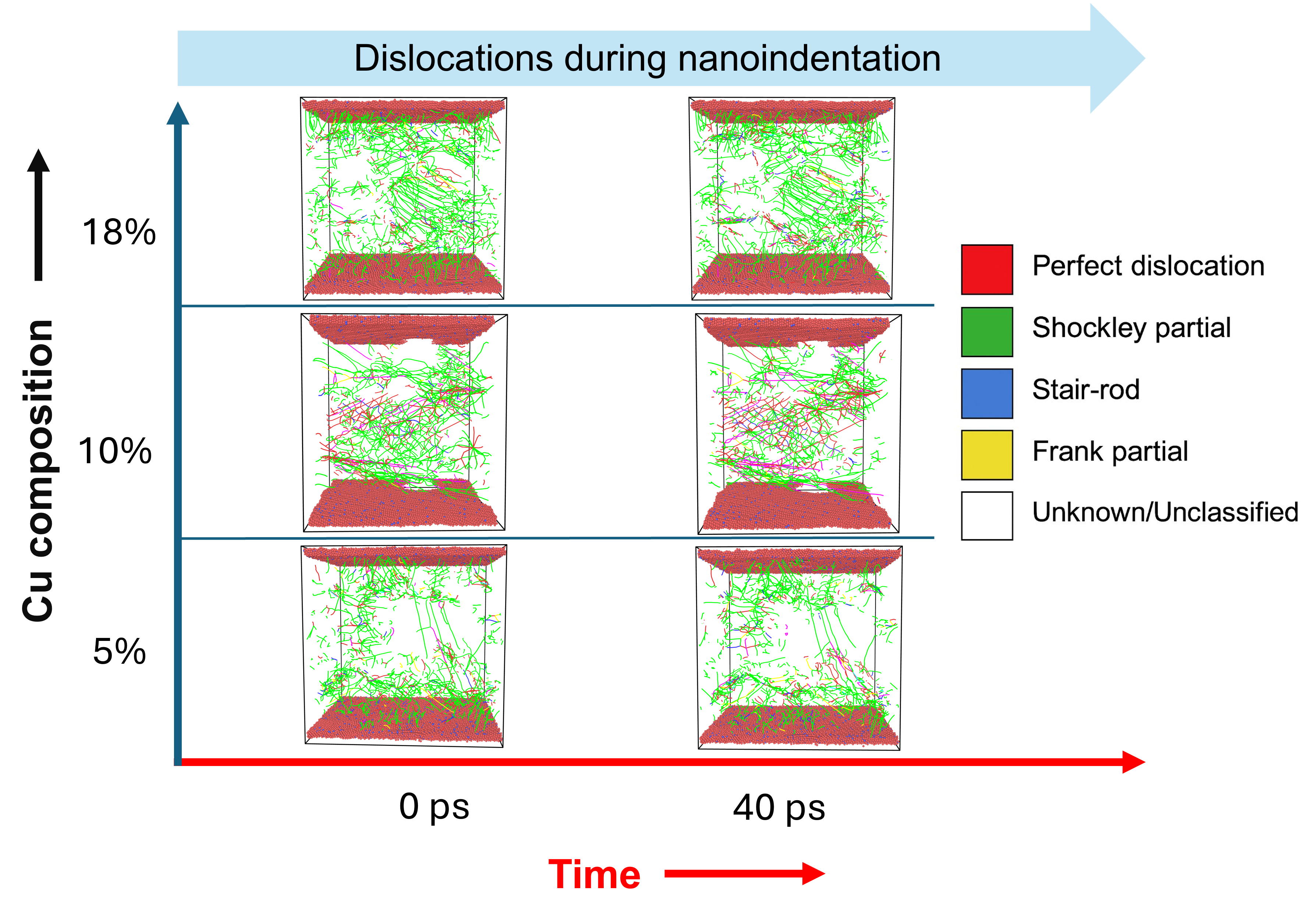}
  \caption{Dislocation analysis of $Al–Cu$ alloys under microgravity at 5 wt.\%, 10 wt.\%, and 18 wt.\% $Cu$. Higher $Cu$ promotes lattice distortions and dislocation nucleation, with 18 wt.\% showing the greatest defect density.}
  \label{fig:14}
\end{figure}

\begin{figure}
  \centering
  \includegraphics[width=0.8\textwidth]{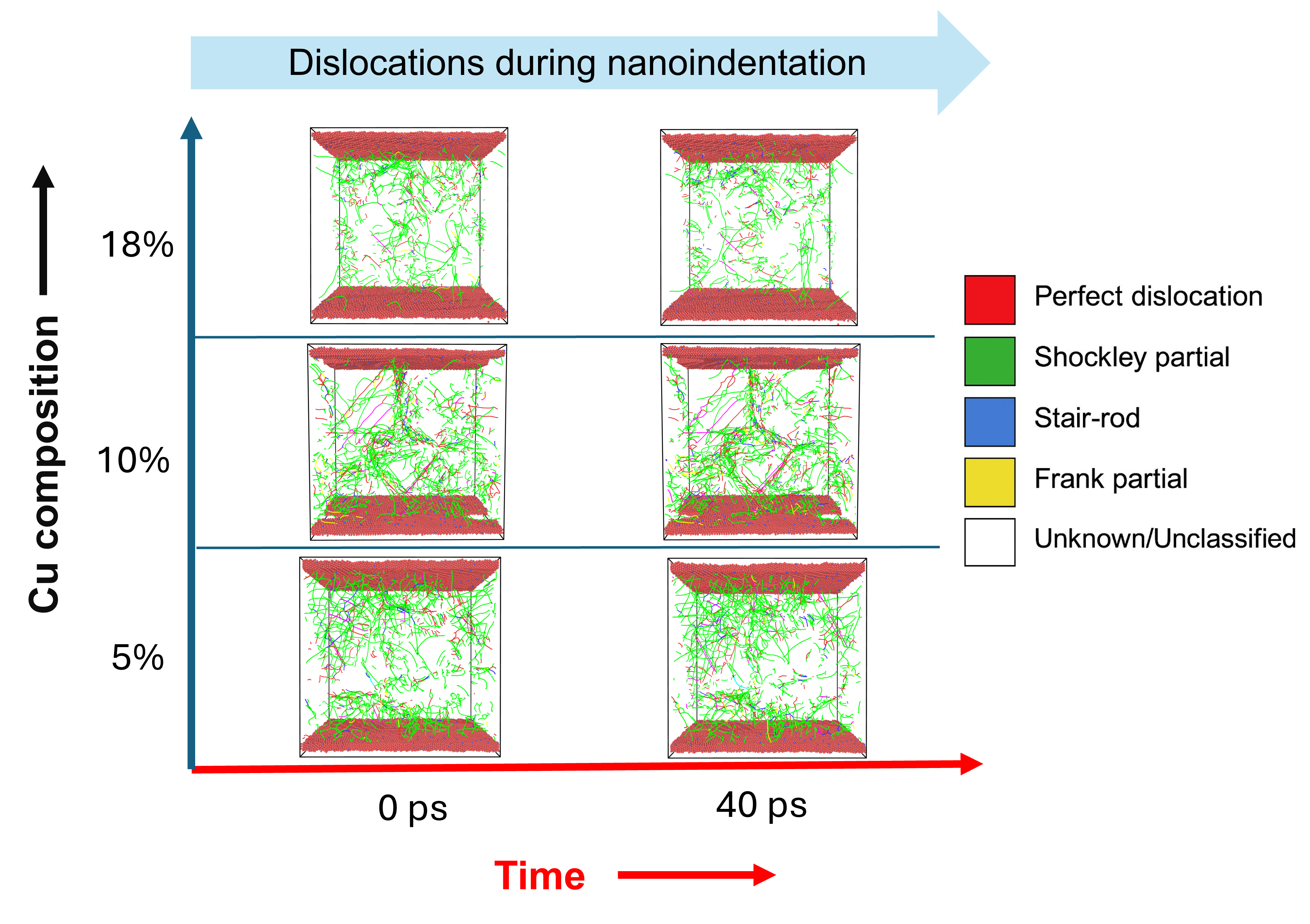}
  \caption{Dislocation analysis of $Al–Cu$ alloys under Earth gravity at 5 wt.\%, 10 wt.\%, and 18 wt.\% $Cu$. Unlike the microgravity case, higher $Cu$ suppresses defect formation, with 5 wt.\% Cu showing the most active dislocation network.}
  \label{fig:15}
\end{figure}

Dislocations are extracted from the nano-indentation simulations, and the analysis is explained in this section. Figures \ref{fig:13}, \ref{fig:14}, and \ref{fig:15} provide direct insight into how dislocation structures evolve under different gravity levels and copper concentrations, thereby clarifying the microscopic origin of the hardness trends.  

Figure \ref{fig:13} shows the dislocation networks in the 5 wt.\% Cu alloy at the start of indentation (0 ps) and after 40 ps for four different gravity conditions. In microgravity, almost no dislocations are observed even at 40 ps, and the lattice remains largely undistorted. As gravity increases to Lunar, Martian, and finally Earth levels, dislocation activity intensifies markedly. At Earth gravity, a dense network of dislocations forms quickly beneath the indenter, showing how gravitational preloading accelerates defect nucleation and growth. This figure establishes the clear role of gravity in supplying an initial defect population: the higher the gravity, the greater the dislocation activity, and thus the stronger the resistance to penetration during nanoindentation.

Figure \ref{fig:14} examines the effect of the composition of alloy under microgravity. Here, the three alloys behave very differently from the gravity-dominated case. The 5 wt.\% and 10 wt.\% systems remain relatively defect-free even after indentation, while the 18 wt.\% alloy develops the highest number of dislocations. Without buoyancy-driven segregation, $Cu$ atoms are distributed uniformly in the melt. At high Cu content, this uniformity enhances local lattice strain and stacking irregularities, which then act as natural nucleation sites for dislocations. As a result, in microgravity, the high-$Cu$ alloy becomes the most defect-rich system, in contrast to the dilute alloys, which maintain nearly perfect FCC structures.  

Figure \ref{fig:15} shows the opposite phenomenon under Earth gravity. In this case, the 5 wt.\% alloy develops the highest dislocation density, while the 10 wt.\% and 18 wt.\% systems exhibit progressively fewer defects. The reason lies in the interaction between gravity and solute segregation. At low $Cu$ content, gravity promotes dislocation nucleation by pulling atoms into stressed regions beneath the indenter. At higher $Cu$ levels, however, gravity enhances segregation at the solid–liquid interface, which stabilizes the lattice and suppresses defect multiplication. Thus, increasing Cu concentration under Earth gravity reduces the overall defect content.  

Comparing Figures \ref{fig:14} and \ref{fig:15} highlights the fundamental inversion between microgravity and Earth gravity. In microgravity, higher $Cu$ promotes dislocation formation through solid-solution lattice distortions, whereas under Earth gravity, higher $Cu$ suppresses dislocations through segregation-induced stabilization. These opposing trends demonstrate that the observed differences in hardness are not simply due to solute content, but instead arise from the way gravity reshapes dislocation generation mechanisms. In other words, the early dislocation content visible in these figures directly determines the indentation resistance measured in the previous subsection: strong gravity strengthens dilute alloys through dislocation multiplication, while microgravity hardens $Cu$-rich alloys through solute-driven lattice distortion in a segregation-free environment.

\begin{figure}
  \centering
  \includegraphics[width=0.65\textwidth]{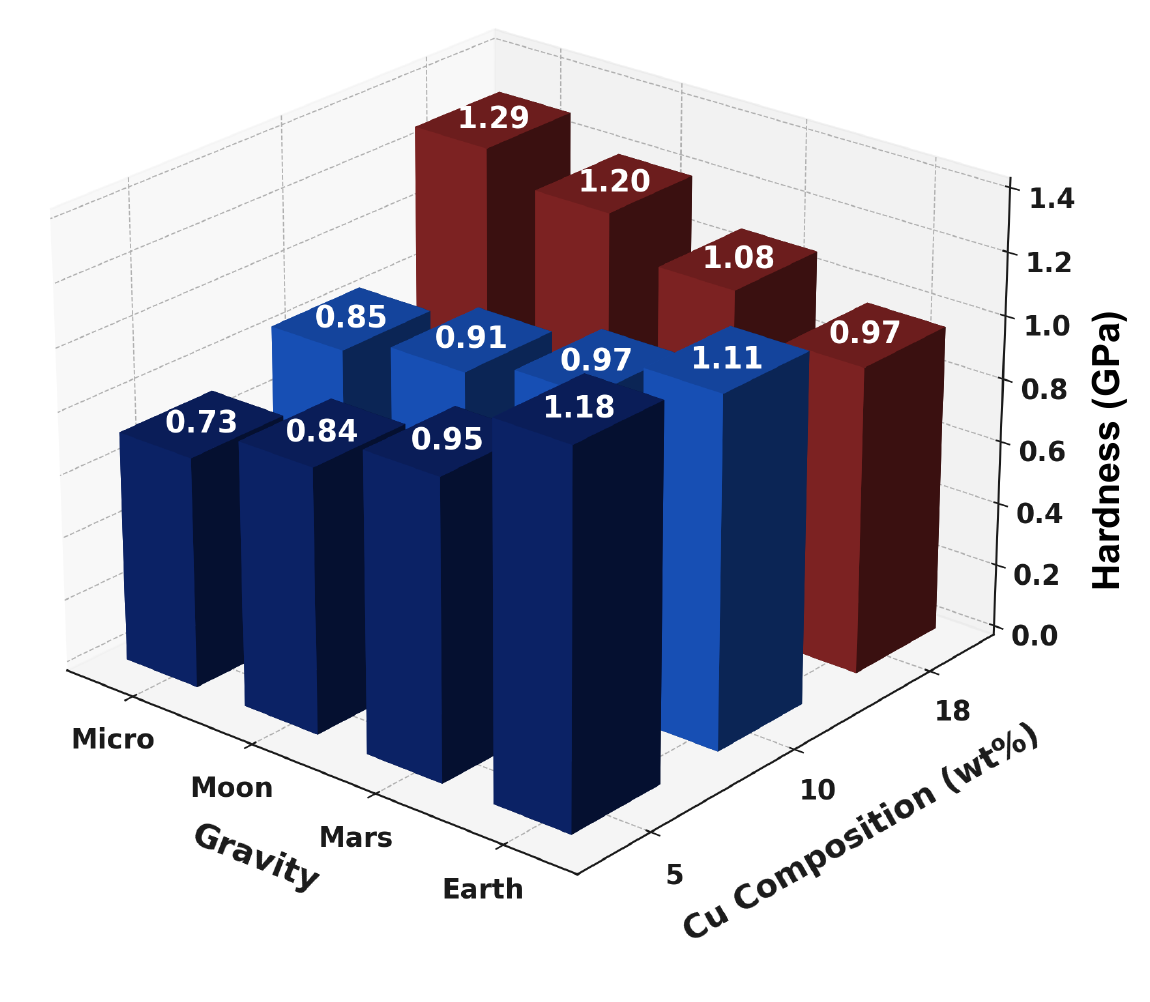}
  \caption{Mechanistic link between hardness and solidification history. Hardness trends reflect the combined influence of gravity-driven defect nucleation and solute-driven strengthening, consistent with dislocation density evolution during solidification.}
  \label{fig:16}
\end{figure}

The hardness values reported in this work are obtained from the nanoindentation simulations by analyzing the indentation force–displacement response. The indenter load is recorded as a function of penetration depth, and the hardness is determined as the ratio of the peak indentation force to the projected contact area, following the standard procedure adapted to atomistic scales. This approach captures the effective resistance of the solid to localized plastic flow and is widely used in both experimental and simulation studies of small-scale mechanical response.

Figure \ref{fig:16} summarizes the variation of hardness as a function of both gravity level and copper composition. At low and intermediate $Cu$ contents (5–10 wt\%), hardness increases with increasing gravity, reaching a maximum under Earth conditions. This trend is physically consistent with the role of hydrostatic compression in suppressing dislocation nucleation beneath the indenter and elevating the stress required for plastic yield. In contrast, the 18 wt\% $Cu$ alloy exhibits the highest hardness under microgravity, which then decreases with increasing gravity. This reversal arises from the effect of solidification history: in microgravity, solute distribution remains uniform and generates significant solid-solution strengthening, while stronger gravity promotes solute segregation and pre-existing dislocation structures that provide easier glide paths for plasticity.

Taken together, these results indicate that the observed hardness trends are governed primarily by the dislocation content and microstructural state inherited from the solidification stage. Low-$Cu$ alloys are hardened mainly by gravity-assisted suppression of dislocation nucleation, while the high-$Cu$ system highlights the competing role of solute distribution and pre-existing defects. This interpretation aligns with known physics of indentation plasticity, where both external pressure and initial defect density dictate the nucleation threshold and subsequent hardness. This outcome supports the hypothesis presented in the earlier part of this article and opens up a new avenue for atomistic investigation on solidification physics outside Earth. 

\section{Limitations and Future Work}
\label{sec:Limitation}

While this study provides new atomistic insights into the coupled influence of $Cu$ percentage and gravity on $Al–Cu$ solidification and nano-indentation response, a few limitations remain. First, the results are derived solely from molecular dynamics simulations without direct experimental validation. Although comparisons with selected benchmark studies suggest that the predicted melting intervals, lattice constants, and qualitative solidification trends are consistent with reported values, no dedicated microgravity solidification or nanoindentation experiments have yet been performed on the exact alloy systems simulated here. The absence of one-to-one validation means that the quantitative values of hardness, dislocation density, and interface morphologies should be interpreted cautiously until corroborated by experiments under controlled gravity conditions.

Second, the interatomic description relies on an EAM potential, which has been widely applied to capture bulk thermodynamics and FCC phase stability. However, this potential does not explicitly stabilize the tetragonal $\theta$ ($Al$$_2$$Cu$) phase, which is a key equilibrium constituent above the eutectic composition. As a result, phase competition between FCC--$Al$ and  $\theta$ ($Al$$_2$$Cu$) cannot be fully represented, and the simulations may underestimate the role of $\theta$-phase formation in influencing solidification pathways and dislocation activity at higher copper contents. Using a more advanced or specifically parameterized $Al-Cu$ potential capable of accurately reproducing $\theta$-phase nucleation and growth would strengthen the fidelity of future simulations.

Finally, the present simulations operate at nanometer length scales and nanosecond time scales. This restricts the accessible cooling rates, microstructural dimensions, and defect evolution compared to laboratory experiments or industrial solidification conditions. While nanoscale insights are valuable for mechanistic understanding, scaling effects must be carefully considered before extrapolating these results to mesoscale or continuum models.

\section{Conclusions}
\label{sec:Conclusion}

In summary, this study establishes an MD framework to investigate how gravity and composition jointly shape the solidification behavior and mechanical properties of $Al$–$Cu$ alloys at the nanoscale. The key findings are:

\begin{itemize}
    \item \textbf{Gravity alters solidification pathways.} Earth and Martian gravity accelerate solidification and promote heterogeneous microstructures with higher dislocation densities, while microgravity delays nucleation and produces more uniform and defect-free crystals.

    \item \textbf{Copper composition couples with gravity.} At low $Cu$ contents (5--10 wt.\%), gravity enhances FCC growth and increases hardness. At high $Cu$ content (18 wt.\%), microgravity stabilizes FCC fronts by suppressing solute segregation, reversing the Earth-based trends.

    \item \textbf{Mechanical response reflects solidification history.} Nanoindentation simulations reveal that hardness trends are consistent with the dislocation content inherited from solidification: higher gravity strengthens dilute alloys by multiplying dislocations, while microgravity hardens $Cu$-rich alloys through uniform solute distribution and solid-solution strengthening.

    \item \textbf{Trends are physically consistent but need further validation.} The investigation aligns qualitatively with known physics of segregation, defect generation, and indentation plasticity. However, quantitative values should be interpreted cautiously until verified by dedicated microgravity experiments, and the inherent limitations of the MD method should be taken into account. 
\end{itemize}

Beyond its immediate insights, this work lays the foundation for predictive modeling of space-based manufacturing processes at the nanoscale. Future efforts will extend this framework to multicomponent alloys, incorporate thermal gradients and interfacial kinetics, and explore integration with mesoscale melt pool simulations. Ultimately, such multiscale modeling could enable intelligent cyberinfrastructure for in-situ alloy design, bridging atomistic mechanisms with mission-ready materials for off-Earth fabrication.

\section{Acknowledgments}
S. Saha gratefully acknowledges the start-up fund provided by the by the Kevin T. Crofton Department of Aerospace and Ocean Engineering, Virginia Polytechnic Institute and State University.

\newpage

\bibliographystyle{elsarticle-num-names} 
\bibliography{refs}





\end{document}